\documentclass[manuscript,screen,nonacm]{acmart} % the settings here are important!
\setcopyright{cc} %% CC BY 4.0 license
\acmConference[]{} %% this used to suppress the adresses footer, but not anymore apparently.

\newcommand{\workshopname}{GenAICHI: CHI 2025 Workshop on Generative AI and HCI}
\newcommand\extrafootertext[1]{% this command adds a non-numbered footnote
    \bgroup
    \renewcommand\thefootnote{\fnsymbol{footnote}}%
    \renewcommand\thempfootnote{\fnsymbol{mpfootnote}}%
    \footnotetext[0]{#1}%
    \egroup
}

\AtBeginDocument{ % setup headers and footers
    \fancypagestyle{firstpagestyle}{
        \fancyhf{}
        \fancyfoot[L]{\sffamily\footnotesize \workshopname}%
        \fancyfoot[C]{\sffamily\footnotesize \thepage}
    }
    \fancyhf{}
    \fancyhead[L]{\sffamily\footnotesize\shorttitle}
    \fancyhead[R]{\sffamily\footnotesize\shortauthors}
    \fancyfoot[L]{\sffamily\footnotesize\workshopname}%
    \fancyfoot[C]{\sffamily\footnotesize\thepage}
}
\usepackage{lipsum} % sample text — remove in production

\begin{document}

%% Title (include an optional short title for running headers)
\title[SakugaFlow]{SakugaFlow: A Stagewise Illustration Framework Emulating the Human Drawing Process and Providing Interactive Tutoring for Novice Drawing Skills}

% -----------------------------------------------------------------
%                          AUTHOR BLOCK (updated with ORCID)
% -----------------------------------------------------------------

% ----------------------------
% Author 1 — Kazuki Kawamura
% ----------------------------
\author{Kazuki Kawamura}
\orcid{0000-0002-5181-320X}
\email{kwmr@acm.org}
\affiliation{%
  \institution{Sony Computer Science Laboratories, Inc.}
  \city{Kyoto}
  \country{Japan}}
\affiliation{%
  \institution{The University of Tokyo}
  \city{Tokyo}
  \country{Japan}}

% ----------------------------
% Author 2 — Jun Rekimoto
% ----------------------------
\author{Jun Rekimoto}
\orcid{0000-0002-3629-2514}
\email{rekimoto@acm.org}
\affiliation{%
  \institution{Sony Computer Science Laboratories, Inc.}
  \city{Kyoto}
  \country{Japan}}
\affiliation{%
  \institution{The University of Tokyo}
  \city{Tokyo}
  \country{Japan}}

% ----------------------------
% Running head configuration
% ----------------------------
\renewcommand{\shortauthors}{Kawamura and Rekimoto}

\begin{abstract} While current AI illustration tools can generate high-quality images from text prompts, they do not explain how a person could replicate the final drawing. We present \textbf{SakugaFlow}, a tool that gradually refines an illustration through four key stages---rough sketch, line art, coloring, and final finishing---and allows users to revise at any point along the way. SakugaFlow integrates diffusion-based image generation at each stage with an AI agent that can advise and consult on how to proceed at the next step. Powered by a Large Language Model, this agent helps novice users understand foundational principles such as anatomy and composition, and offers localized revision guidance. By revealing intermediate outputs at each stage, our system transforms generative AI from a black-box producer into a \emph{learning partner}, enabling beginners to develop their skills as they iteratively refine their artwork. \end{abstract}




\begin{teaserfigure}
  \centering
  \includegraphics[width=\textwidth]{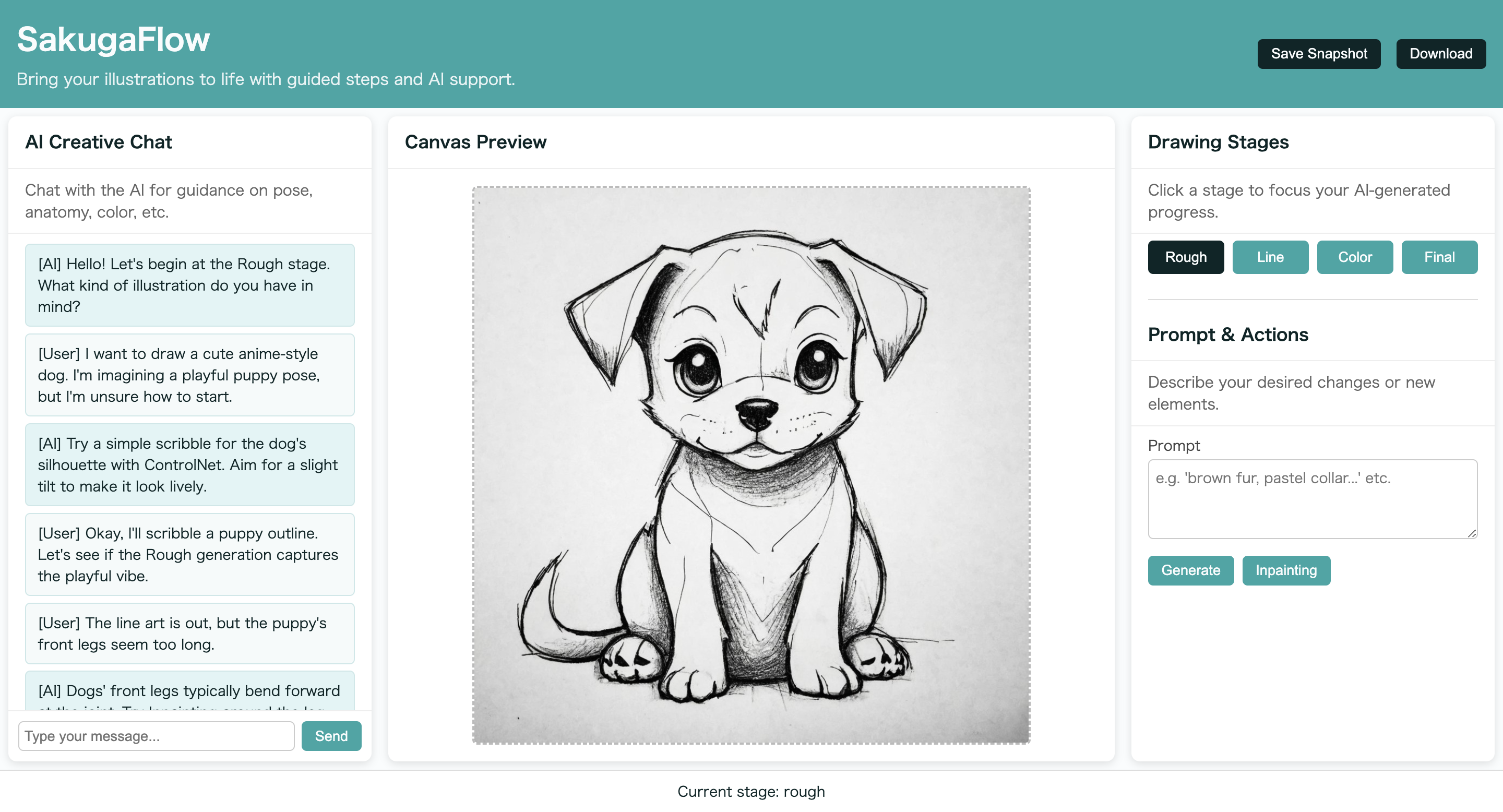}
  \vspace{-3em}
  \caption{Overview of SakugaFlow's UI and architecture. (1) Users progress from rough to line to color to finish, each powered by a diffusion-based backend. (2) An LLM tutor provides real-time feedback on anatomy, perspective, and more. (3) Users can branch versions, inpaint local regions, and compare alternatives within the same project.}
  \Description{Conceptual diagram showing the main UI flow and system architecture for SakugaFlow.}
  \label{fig:teaser}
\end{teaserfigure}


\maketitle

\section{Introduction}
\vspace{-0.3em}
Generative models for illustration have advanced rapidly, with diffusion-based approaches now producing remarkably detailed images from simple prompts \cite{Ho2020,Rombach_2022_CVPR}. However, these one-shot techniques rarely show the \emph{drawing process} that underlies the final output. For a beginner hoping to improve their own skills, seeing only a completed image provides little guidance on \emph{how} lines, colors, or forms evolve.

Human artists, by contrast, usually begin with a loose rough sketch, transition to cleaner line art, apply broad color blocks, and then refine shading and details. Each step can be a learning opportunity: understanding silhouette, proportion, color harmony, and so forth. If an AI system bypasses—or at least does not reveal—these steps, novices miss the gradual refinement process that connects simple fundamentals to complex finished pieces.

SakugaFlow addresses this gap. Rather than presenting an immediate final result, our system divides the illustration pipeline into four explicit stages: rough sketch, line art, coloring, and final finishing. At each stage, users see a partially complete image generated by a diffusion model, which they can revise or question with the help of an educational dialogue agent. This agent, built on a Large Language Model (LLM), offers targeted explanations about anatomy, perspective, and design principles. By directly associating each incremental change with an interpretable step, SakugaFlow transforms a black-box generator into a \emph{scaffolded learning environment}, where users not only observe but also \emph{actively influence} how the image develops.

\noindent \textbf{Contributions.} Our key contributions are as follows: 
\begin{itemize}
\item \textbf{Staged diffusion aligned with human drawing steps.} While prior diffusion methods mainly offer localized control of final outputs, we re-map them into four phases: rough sketch, line art, coloring, and finishing. This approach clarifies \emph{how} the final image emerges and allows novices to experiment with each step. 
\item \textbf{Integration of an educational dialogue agent for real-time feedback.} Although interactive tutoring systems have seen success in text-based domains, their application to drawing is underexplored. Our system incorporates an LLM to connect each generative step with relevant art principles, offering immediate explanations (e.g., perspective corrections) and adaptive guidance throughout the process. 
\item \textbf{Fusing creative assistance with skill development.} Unlike prior tools focused on final output or user creativity, we emphasize helping beginners build foundational illustration skills. By positioning AI as both teacher and partner, we enable stepwise practice, explorative feedback, and reflective learning—prioritizing skill acquisition over mere production efficiency. \end{itemize}
\vspace{-0.5em}

%%%%%%%%%%%%%%%%%%%%%%%%%%%%%%%%%%%%%%%%%%%%%%%%%%%%%%%%%%%%
\section{Related Work}
\vspace{-0.3em}
\noindent\textbf{AI-based Illustration Generation: Focus on Diffusion Models.}
% 拡散モデル~~Ho2020,Rombach_2022_CVPR}は、GAN~~Goodfellow2014, radford2015unsupervised, zhu2017unpaired, karras2019style}の強力な代替手段として注目を集めており、反復的なノイズ除去によって高品質な画像生成を可能にしている。ControlNet、Prompt-to-Prompt、Inpainting~~Zhang_2023_ICCV,hertz2022prompt}のような拡張は、線画、ポーズ指定、テキストベースの編集に対する洗練された制御を可能にし、Promptify~~Brade2023Promptify}も同様に、大規模な言語モデルを使用したインタラクティブなプロンプト探索を提供する。しかし、これらの既存の手法は、主に完成画像の作成または編集に焦点を当てており、人間がその画像をどのように描くかという描画プロセスをモデル化しておらず、そのような情報を抽出することもできない。
Diffusion models~\cite{Ho2020,Rombach_2022_CVPR} have gained prominence as a powerful alternative to GANs~\cite{Goodfellow2014, radford2015unsupervised, zhu2017unpaired, karras2019style}, enabling high-quality image generation through iterative noise removal. Extensions like ControlNet, Prompt-to-Prompt, and Inpainting~\cite{Zhang_2023_ICCV,hertz2022prompt} allow refined control over line art, pose specification, and text-based editing, while Promptify~\cite{Brade2023Promptify} similarly offers interactive prompt exploration using LLM. However, these existing methods focus primarily on the creation or editing of the finished image, and do not model the drawing process of how humans complete their drawing from nothing, nor are they able to extract such information.

\noindent\textbf{Illustration Learning Support and Process Visualization.}
Tutorial videos and step-by-step guides are widely available, but they often end up being passive viewing, and it is difficult to choose the learning material you want. Logging or replay of drawing processes~\cite{rubaiat2010chronicle,Jennifer2011} can show expert workflows via timelapse but typically lack real-time user input for revisions or questions. Real-time drawing aids~\cite{Blake2017} focus on hand-drawn input, such as automatic error correction or feedback, but have not yet explored the perspective of \emph{learning from AI-generated} intermediate outputs.

\noindent\textbf{Creative Support and Human-AI Co-Creation.}
\noindent
Recent commercial software, such as Photoshop (Generative Fill) and Illustrator (Generative Recolor), integrates AI for quick image and vector transformations~\cite{swift2024}.
Academic studies likewise investigate real-time AI collaboration in drawing~\cite{davis2025,Nicholas2016} and prompt-based image generation~\cite{Brade2023Promptify,Wang2024}.
While many of these systems boost final-image efficiency or assist with ideation, they rarely address how novices \emph{learn} illustration step by step.
We propose a stage-based AI approach combined with educational dialogues, allowing beginners to observe each process phase---rough sketch, line art, coloring, and finalizing---and receive guided support.
By focusing on progressive skill acquisition rather than just output optimization, our work fills an important gap in existing co-creation research.

%%%%%%%%%%%%%%%%%%%%%%%%%%%%%%%%%%%%%%%%%%%%%%%%%%%%%%%%%%%%
\section{Proposed System: SakugaFlow}
\label{sec:sakugaflow}

\vspace{-0.3em}
\subsection{Design Goals and Rationale}
\vspace{-0.3em}
\noindent
SakugaFlow aims to help beginners learn how illustrations progress through key stages rather than jumping directly to a final image. Human artists typically do a rough sketch, refine line art, block colors, then add finishing details. We integrate a diffusion model into these steps so novices can see partial outputs and adjust them step by step. This structure ties closely to educational needs, providing a guided, interpretable journey through each stage of the creative process.

\vspace{-1.0em}
\subsection{System Overview and UI}
\vspace{-0.3em}
\noindent
As shown in Fig.\ref{fig:teaser}, SakugaFlow presents a four-phase workflow:
\begin{enumerate}
    \item \textbf{Rough Sketch:} Users define basic shapes or a scribble-based layout (optionally with ControlNet).
    \item \textbf{Line Art:} A Prompt-to-Prompt transformation clarifies contours; Inpainting can refine errors (e.g., proportions).
    \item \textbf{Coloring:} The system suggests color palettes or shading styles; users can branch multiple color studies.
    \item \textbf{Finishing:} Final highlights, lighting, and minor adjustments lead to the completed illustration.
\end{enumerate}

A chat pane (right or left side) connects to an LLM-based “tutor,” which offers quick theory tips (e.g., “Why add line thickness here?”). Users can also request partial re-generation by selecting areas to inpaint or rewriting the prompt. A branching manager lets them save and compare multiple versions of the same stage.

\vspace{-1.0em}
\subsection{Interaction Flow}
\vspace{-0.3em}
\noindent
\textbf{1) Project Setup.} The user inputs a theme (e.g., “fantasy character”), then starts at the Rough stage.  
\textbf{2) Rough Stage.} ControlNet(\emph{Scribble}) or a blank canvas yields a rough shape. The LLM tutor might suggest adjusting pose or composition.  
\textbf{3) Line Art Stage.} Prompt-to-Prompt refines the lines; users correct details (e.g., mismatched arm length) via Inpainting.  
\textbf{4) Color Stage.} The system proposes palettes; a user can branch to compare two coloring approaches. The tutor explains color theory (e.g., warm vs.\ cool contrast).  
\textbf{5) Finishing.} Final touches (lighting, shading). The tutor checks consistency and prompts reflection (“Where is the light source?”).  

Users can backtrack or create alternative branches. \emph{Active experimentation} is encouraged rather than showing a single final image.

\vspace{-1.0em}
\subsection{Implementation Details}
\vspace{-0.3em}
\noindent
\textbf{Frontend.} Built in React with a canvas for partial previews, an Inpainting brush, and a chat interface. Minimizing latency is crucial; we use a small in-browser aggregator for partial updates.  
\textbf{Backend.} Python (FastAPI) orchestrates:
\begin{itemize}
    \item \textit{Diffusion Model:} Stable Diffusion + ControlNet for each stage.  
    \item \textit{LLM (GPT):} Processes user queries and context from the current stage to generate relevant tips.  
    \item \textit{Versioning/Branches:} Stores user steps, prompts, and image states, enabling revert and compare.  
\end{itemize}
All major computations run on a GPU server, returning images to the frontend asynchronously. Future optimizations include caching partial results for repeated local edits.

\vspace{-1.0em}
\subsection{Optional Hand-Drawn Input}
\vspace{-0.3em}
\noindent
While novices can rely solely on text prompts and partial scribbles, advanced users may integrate their own detailed sketches to preserve personal style or practice motor skills. SakugaFlow can treat these sketches like “control images,” merging them with generated content.

\vspace{-0.3em}

\begin{figure}[t]
 \centering
 \includegraphics[width=0.98\linewidth]{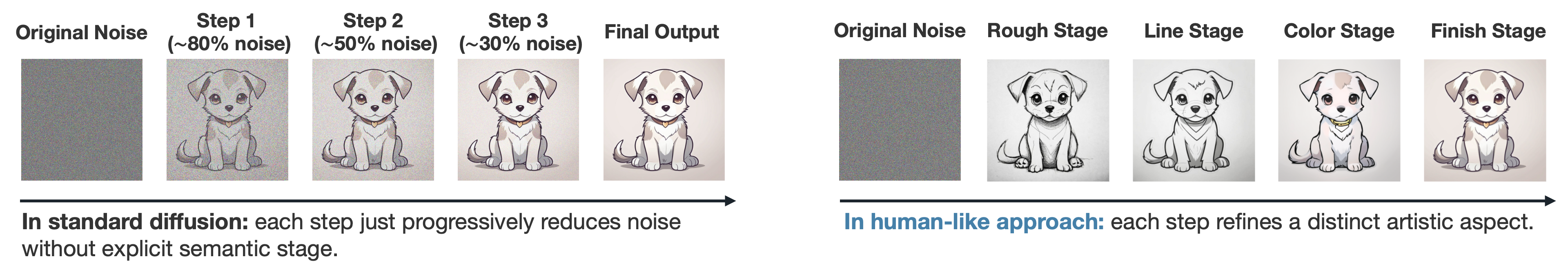}
 \vspace{-1em}
  \caption{
  \textbf{A human-like stepwise approach (left) vs.\ a standard diffusion process (right).}
  Top: the illustration evolves through rough, line, color, and final stages. 
  Bottom: the model gradually removes noise (e.g., $\sim80\%\!\rightarrow\!0\%$) without clear drawing phases.
  }
 \label{fig:diffusion_step}
\end{figure}

%%%%%%%%%%%%%%%%%%%%%%%%%%%%%%%%%%%%%%%%%%%%%%%%%%%%%%%%%%%%
\section{Discussion \& Future Work}
\noindent \textbf{Limitations and Future Directions.}
Although SakugaFlow introduces separate steps (rough, line, color, finish), our backend diffusion model remains optimized for final outputs, limiting the interpretability of intermediate states. Future efforts include constructing more granular representations (e.g., partial sketches or layered data), training explicitly on sequential refinements, and running controlled user studies to quantify skill acquisition. Balancing fast, one-shot generation with slower, staged processes is also crucial for scenarios where iterative learning is desired.

\noindent \textbf{Toward a Human-Like Staged Diffusion Approach}
While standard diffusion treats image generation as a continuous noise-removal sequence, a human-like staged approach aligns more closely with how artists actually work. Fig.~\ref{fig:comparison} contrasts these methods: the top row shows discrete phases (rough, line, color, final), while the bottom row removes noise without revealing clear drawing steps. By making each phase explicit, we can offer finer-grained visualization of \emph{when} lines, color, and details emerge, providing a more intuitive interaction for novices who naturally think, ``draw lines first, then add color, then refine.'' Meanwhile, these intermediate stages can be reused as modular sub-tasks (e.g., silhouette extraction, line analysis, color application), opening the door to new multi-task pipelines or partial retraining. Finally, segmenting the process in a human-like manner enhances interpretability and debugging: if a line stage is flawed, we know exactly where to intervene. This also helps in settings where safety or quality assurance requires each step to be validated before proceeding.

In conclusion, SakugaFlow represents an initial step toward interpretable, pedagogically rich generative illustration. Further aligning the underlying models with human workflows—and rigorously evaluating those alignment strategies—remains our primary goal for future work.

\bibliographystyle{ACM-Reference-Format}
\bibliography{reference}

\end{document}